\documentclass{revtex4}
\usepackage{amsfonts}
\usepackage{amsmath}

\input{tcilatex}

\begin{document}

\title{Double-bounce domain-wall in Einstein-Yang-Mills-Scalar black holes}
\author{S. Habib Mazharimousavi}
\author{M. Halilsoy}
\author{T. Tahamtan}
\email{habib.mazhari@emu.edu.tr\\
mustafa.halilsoy@emu.edu.tr \\
tayabeh.tahamtan@emu.edu.tr}

\begin{abstract}
We find Einstein-Yang-Mills (EYM) black hole solutions endowed with massless
scalar hair in the presence of a potential $V\left( \phi \right) $ as
function of the scalar field $\phi $. Choosing $V\left( \phi \right) =$%
constant (or zero) sets the scalar field to vanish leaving us with the EYM
black holes. Our class of black hole solutions is new so that they do not
asymptote in general to any known limits. Particular case is given, however,
which admits an asymptotically anti de Sitter limit in $6-$dimensional
spacetime. The role of the potential $V\left( \phi \right) $ in making
double bounces (i.e. both a minimum and maximum radii) on a Domain Wall (DW)
universe is highlighted.
\end{abstract}

\maketitle

\section{\ \ \ \ \ \noindent Introduction}

There has always been curiosity in obtaining exact solutions in Einstein's
theory that contain new sources to encompass the underlying spacetime
curvature. This covers sources such as scalar (charged, uncharged),
cosmological constant, electromagnetism (linear, non-linear, electric,
magnetic), Yang-Mills (YM) fields plus others as well as their combinations.
The necessity of enrichment in sources can be traced back to gauge / gravity
duality, but may find better justification from the more recent holographic
superconductivity analogy. Once spherical symmetry and asymptotic flatness
in the spacetime ansatz are assumed severe restrictions on possible
solutions are inevitable. The uniqueness theorems and no-hair conjectures
are a few of such restrictions that the solutions obtained must comply with.

In a recent study minimally coupled scalar field has been considered in
Einstein-Maxwell theory with anti-de Sitter (AdS) asymptotics {\cite{1}-a
(More recently, the same authors studied also the static, }planar solutions
of Einstein-scalar gravity which results an anti-de Sitter vacuum \cite{1}%
-b).

In \cite{1} beside kinetic Lagrangian terms due to the scalar and Maxwell
fields an additional potential $V(\phi )$ constructed from scalar field has
been supplemented. Constancy of such a potential can naturally be
interpreted as a cosmological term. In some specific cases the scalar field
is assumed to be of the form $\phi \sim \ln r$ and upon this choice the
potential $V(\phi )$ can be determined from the dynamical equations. Among
other things the model is shown to admit asymptotic Lifshitz black holes.

In this paper we add Yang-Mills (YM) field instead of Maxwell to the
Einstein-minimally coupled scalar field system and search for the resulting
solutions. Our pure magnetic YM field is added through the Wu-Yang ansatz
which was introduced before {\cite{2,3}}, and is known to extend easily to
all higher dimensions. Our model of Einstein-Scalar-YM system contains an
indispensable potential function $V(\phi )$ as a function of the scalar
field. With the exception of specific dimensions and scalar field our
solutions do not admit immediately recognizable asymptotes. In $6-$%
dimensions, for instance we obtain an EYM-scalar solution which asymptotes
to an anti-de Sitter spacetime. Although spherical symmetry allows more
general ansatzes our choice in this paper will be such that $-g_{tt}=g^{rr}$%
, while the coefficient of the unit angular line element $d\Omega _{d}^{2}$
is an arbitrary function $R(r)^{2}$. The asymptotic flatness condition
requires that $R(r)=r$, which should be discarded in the present case since
it annules the scalar field through the field equations. We observe also
that non-trivial scalar and YM fields must coexist only by virtue of a
non-zero potential $V(\phi )$. The choice $V(\phi )=$constant (or zero)
admits no scalar solution. Overall, our solutions can be interpreted as
hairy black holes in an asymptotically non-flat spacetime. By setting the
scalar field to zero we recover the EYM black holes obtained before {\cite{3}%
}. Our model is exemplified with specific parameters of scalar hair. Next,
in $d+2-$dimensional bulk spacetime we define a $d+1-$dimensional
Domain-Wall (DW) as a Friedmann-Robertson-Walker (FRW) universe by the
proper boundary (junction) conditions given long ago by Darmois and Israel {%
\cite{4}}. We establish such a DW universe and explore the conditions under
which the FRW universe has double bounces. That is, the radius function of
FRW universe will lie in between the minimum and maximum values. We had
shown previously the existence of such double bounces in different theories {%
\cite{5,6}}.

The paper is organized as follows. Our formalism of EYM system with a scalar
field and potential is introduced in Section II. Section III presents exact
solution to the system under certain ansatzes. Section IV gives a solution
with a different scalar field ansatz in 6-dimensional spacetime that
asymptots to anti de Sitter spacetime. Domain wall dynamics in our bulk
space is introduced in Section V. Our conclusion appears in Section VI.

\section{The formalism}

Following the formalism given in {\cite{1}} with the same unit convention
(i.e., $c=16\pi G=1$) the $d+2-$dimensional Einstein-Yang-Mills gravity
coupled minimally to a scalar field $\phi $ is given by 
\begin{equation}
S=\int d^{d+2}x\sqrt{-g}\left[ \mathcal{R}-2\left( \partial \phi \right)
^{2}-L_{\left( YM\right) }-V\left( \phi \right) \right]
\end{equation}%
in which $\mathcal{R}$ is the Ricci scalar, $L_{\left( YM\right) }=Tr\left(
F_{\mu \nu }^{\left( a\right) }F^{\left( a\right) \mu \nu }\right) $ and $%
V\left( \phi \right) $ is an arbitrary function of the scalar field $\phi $.
The YM field $2-$form components are given by 
\begin{equation}
\mathbf{F}^{\left( a\right) }=\frac{1}{2}F_{\mu \nu }^{\left( a\right)
}dx^{\mu }\wedge dx^{\nu }
\end{equation}%
with the internal index $(a)$ running over the degrees of freedom of the
nonabelian YM gauge field. Variation of the action with respect to the
metric $g_{\mu \nu }$ gives the EYM field equations as 
\begin{equation}
G_{\mu }^{\nu }=T_{\mu }^{\nu \left( YM\right) }+T_{\mu }^{\nu \left( \phi
\right) }-\frac{1}{2}V\left( \phi \right) \delta _{\mu }^{\nu }
\end{equation}%
in which 
\begin{equation}
T_{\mu }^{\nu \left( YM\right) }=2Tr\left( F_{\mu \alpha }^{\left( a\right)
}F^{\left( a\right) \nu \alpha }-\frac{1}{4}F_{\gamma \sigma }^{\left(
a\right) }F^{\left( a\right) \gamma \sigma }\delta _{\mu }^{\nu }\right) ,%
\text{ }T_{\mu }^{\nu \left( \phi \right) }=2\left( \partial _{\mu }\phi
\partial ^{\nu }\phi -\frac{1}{2}\partial _{\rho }\phi \partial ^{\rho }\phi
\delta _{\mu }^{\nu }\right) .
\end{equation}

Variation of the action with respect to the scalar field $\phi $ yields 
\begin{equation}
\nabla ^{2}\phi =\frac{1}{4}\frac{dV\left( \phi \right) }{d\phi }.
\end{equation}%
The $SO\left( d+1\right) $ gauge group YM potentials are given by {\cite{7}} 
\begin{equation}
\begin{array}{ll}
\mathbf{A}^{(a)} & =\frac{Q}{r^{2}}C_{\left( i\right) \left( j\right)
}^{\left( a\right) }x^{i}dx^{j},\text{{}}Q=\text{$\func{YM}$ $\func{magnetic}
$ $\func{charge}$,}r^{2}=\sum_{i=1}^{d+1}x_{i}^{2}, \\ 
2 & \leq j+1\leq i\leq d+1,\text{$\func{and}$}1\leq a\leq d\left( d+1\right)
/2, \\ 
x_{1} & =r\cos \theta _{d-1}\sin \theta _{d-2}...\sin \theta
_{1},x_{2}=r\sin \theta _{d-1}\sin \theta _{d-2}...\sin \theta _{1}, \\ 
x_{3} & =r\cos \theta _{d-2}\sin \theta _{d-3}...\sin \theta
_{1},x_{4}=r\sin \theta _{d-2}\sin \theta _{d-3}...\sin \theta _{1}, \\ 
& ... \\ 
x_{d} & =r\cos \theta _{1},%
\end{array}%
\end{equation}%
in which $C_{\left( b\right) \left( c\right) }^{\left( a\right) }$ are the
non-zero structure constants of $\frac{d\left( d+1\right) }{2}-$parameter
Lie group $\mathcal{G}$ {\cite{2,3}}. The metric ansatz is spherically
symmetric which reads 
\begin{equation}
ds^{2}=-U\left( r\right) dt^{2}+\frac{dr^{2}}{U\left( r\right) }+R\left(
r\right) ^{2}d\Omega _{d}^{2},
\end{equation}%
with the only unknown functions $U\left( r\right) $ and $R\left( r\right) $
and the solid angle element 
\begin{equation}
d\Omega _{d}^{2}=d\theta _{1}^{2}+\sum_{i=2}^{d}\prod_{j=1}^{i-1}\sin
^{2}\theta _{j}\hspace{0.75em}d\theta _{i}^{2},
\end{equation}%
with 
\begin{equation*}
0\leq \theta _{d}\leq 2\pi ,0\leq \theta _{i}\leq \pi ,\text{{}}1\leq i\leq
d-1.
\end{equation*}%
Variation of the action with respect to $\mathbf{A}^{\left( a\right) }$
implies the YM equations 
\begin{equation}
\mathbf{d}^{\star }\mathbf{F}^{\left( a\right) }+\frac{1}{\sigma }C_{\left(
b\right) \left( c\right) }^{\left( a\right) }\mathbf{A}^{\left( b\right)
}\wedge ^{\star }\mathbf{F}^{\left( c\right) }=0,
\end{equation}%
in which $\sigma $ is a coupling constant and "$^{\star }$\textbf{"} means
duality. The YM invariant is given by 
\begin{equation}
Tr\left( F_{\mu \nu }^{\left( a\right) }F^{\left( a\right) \mu \nu }\right) =%
\frac{d\left( d-1\right) Q^{2}}{r^{4}}
\end{equation}%
and 
\begin{equation}
Tr\left( F_{t\alpha }^{\left( a\right) }F^{\left( a\right) t\alpha }\right)
=Tr\left( F_{r\alpha }^{\left( a\right) }F^{\left( a\right) r\alpha }\right)
=0,
\end{equation}%
while 
\begin{equation}
Tr\left( F_{\theta _{i}\alpha }^{\left( a\right) }F^{\left( a\right) \theta
_{i}\alpha }\right) =\frac{\left( d-1\right) Q^{2}}{r^{4}},
\end{equation}%
which leads us to the closed form of the energy momentum tensor 
\begin{equation}
T_{\mu }^{\nu \left( YM\right) }=-\frac{d\left( d-1\right) Q^{2}}{2R^{4}}%
\text{$\func{diag}$}\left[ 1,1,\frac{\left( d-4\right) }{d},\frac{\left(
d-4\right) }{d},...,\frac{\left( d-4\right) }{d}\right] .
\end{equation}%
The above field equations can be rearranged as 
\begin{equation}
\frac{R^{\prime \prime }}{R}=-\frac{2}{d}\left( \phi ^{\prime }\right) ^{2},%
\text{{}}\left( UR^{d}\phi ^{\prime }\right) ^{\prime }=\frac{1}{4}R^{d}%
\frac{dV}{d\phi },
\end{equation}%
\begin{equation}
\left( UR^{d}\right) ^{\prime \prime }=d\left( d-1\right)
R^{d-2}+2R^{d}\left( T_{\theta _{i}}^{\theta _{i}\left( YM\right) }+\frac{2}{%
d}T_{r}^{r\left( YM\right) }\right) -\frac{d+2}{d}R^{d}V,
\end{equation}%
\begin{equation}
\left( R^{d-1}UR^{\prime }\right) ^{\prime }=\left( d-1\right) R^{d-2}+\frac{%
2}{d}R^{d}T_{r}^{r\left( YM\right) }-\frac{1}{d}R^{d}V,
\end{equation}%
in which a 'prime' denotes derivative with respect to $r$. As it was
introduced in Ref. {\cite{1}} we define new variables 
\begin{equation}
F\left( r\right) =-\frac{2}{d}\left( \phi ^{\prime }\right) ^{2},R=e^{\int
Ydr},\text{{}}u=UR^{d},
\end{equation}%
which reduce the field equations into 
\begin{equation}
Y^{\prime }+Y^{2}=F\left( r\right) ,\text{{}}\left( u\phi ^{\prime }\right)
^{\prime }=\frac{1}{4}e^{d\int Ydr}\frac{dV}{d\phi },
\end{equation}%
\begin{equation}
u^{\prime \prime }=d\left( d-1\right) e^{\left( d-2\right) \int
Ydr}+2e^{d\int Ydr}\left( T_{\theta _{i}}^{\theta _{i}\left( YM\right) }+%
\frac{2}{d}T_{r}^{r\left( YM\right) }\right) -\frac{d+2}{d}e^{d\int Ydr}V,
\end{equation}%
\begin{equation}
\left( uY\right) ^{\prime }=\left( d-1\right) e^{\left( d-2\right) \int Ydr}+%
\frac{2}{d}e^{d\int Ydr}T_{r}^{r\left( YM\right) }-\frac{1}{d}e^{d\int Ydr}V.
\end{equation}%
A combination of the last two equations yields 
\begin{equation}
u^{\prime \prime }-\left( d+2\right) \left( uY\right) ^{\prime }=-2\left(
d-1\right) e^{\left( d-2\right) \int Ydr}+2e^{d\int Ydr}\left( T_{\theta
_{i}}^{\theta _{i}\left( YM\right) }-T_{r}^{r\left( YM\right) }\right)
\end{equation}%
which is a $V-$independent equation and can be integrated once to 
\begin{equation}
u^{\prime }-\left( d+2\right) uY=\int \left\{ -2\left( d-1\right) e^{\left(
d-2\right) \int Ydr}+2e^{d\int Ydr}\left( T_{\theta _{i}}^{\theta _{i}\left(
YM\right) }-T_{r}^{r\left( YM\right) }\right) \right\} dr+C_{1}.
\end{equation}%
This is further integrated to obtain 
\begin{equation}
u=R^{d+2}\left[ \int \left( \frac{1}{R^{d+2}}\left[ \int \left\{ -2\left(
d-1\right) R^{d-2}+2R^{d}\left( T_{\theta _{i}}^{\theta _{i}\left( YM\right)
}-T_{r}^{r\left( YM\right) }\right) \right\} dr+C_{1}\right] \right) dr+C_{2}%
\right]
\end{equation}%
and consequently the potential reads out from (19) as 
\begin{equation}
V=\frac{d^{2}\left( d-1\right) }{\left( d+2\right) R^{2}}+\frac{2d}{\left(
d+2\right) }\left( T_{\theta _{i}}^{\theta _{i}\left( YM\right) }+\frac{2}{d}%
T_{r}^{r\left( YM\right) }\right) -\frac{d}{\left( d+2\right) R^{d}}%
u^{\prime \prime }.
\end{equation}

\section{Exact Solutions}

We start with an ansatz for the scalar field $\phi =\alpha \ln \left( \frac{r%
}{r_{0}}\right) $ in which $r_{0}$ and $\alpha $ are two real constants. The 
$Y-$equation then, takes the Riccati form 
\begin{equation}
Y^{\prime }+Y^{2}=-\frac{2}{d}\frac{\alpha ^{2}}{r^{2}},
\end{equation}%
which admits a solution for $Y$ given by 
\begin{equation}
Y=\frac{A}{r}
\end{equation}%
with 
\begin{equation}
\begin{array}{l}
A^{2}-A+\frac{2\alpha ^{2}}{d}=0 \\ 
0<A<1.%
\end{array}%
\end{equation}%
Note that $A=0,1$ make the scalar field to vanish, so we exclude them.
Knowing $Y$ we find 
\begin{equation}
R=\left( \frac{r}{r_{1}}\right) ^{A},\text{(}r_{1}=cons.\text{)}
\end{equation}%
and therefore 
\begin{equation}
u=r^{A\left( d+2\right) }\left[ \int \left( \frac{\int \left[ -2\left(
d-1\right) r^{A\left( d-2\right) }r_{1}^{-A\left( d-2\right) }+4\left(
d-1\right) Q^{2}r^{A\left( d-4\right) }r_{1}^{-A\left( d-4\right) }\right]
dr+C_{1}}{r^{A\left( d+2\right) }}\right) dr+\frac{C_{2}}{r_{1}^{A\left(
d+2\right) }}\right]
\end{equation}%
which becomes 
\begin{equation}
u=\left\{ 
\begin{array}{lcr}
\frac{\left( d-1\right) r^{2+A\left( d-2\right) }r_{1}^{-A\left( d-2\right) }%
}{\left( A\left( d-2\right) +1\right) \left( 2A-1\right) }-\frac{2\left(
d-1\right) Q^{2}r^{A\left( d-4\right) +2}r_{1}^{-A\left( d-4\right) }}{%
\left( A\left( d-4\right) +1\right) \left( 3A-1\right) }-\frac{C_{1}}{%
A\left( d+2\right) -1}r+C_{2}\left( \frac{r}{r_{1}}\right) ^{A\left(
d+2\right) } & , & A\neq \frac{1}{2},\frac{1}{3},\frac{1}{d+2} \\ 
-\frac{4\left( d-1\right) \ln (r)r_{1}^{-\frac{d-2}{2}}r^{\frac{d+2}{2}}}{d}-%
\frac{8\left( d-1\right) Q^{2}r^{\frac{d}{2}}r_{1}^{-\frac{d-4}{2}}}{d-2}-%
\frac{2C_{1}}{d}r+C_{2}\left( \frac{r}{r_{1}}\right) ^{\frac{\left(
d+2\right) }{2}} & , & A=\frac{1}{2} \\ 
-\frac{9\left( d-1\right) r_{1}^{\frac{-d+2}{3}}r^{\frac{d+4}{3}}}{\left(
d+1\right) }+12Q^{2}r_{1}^{\frac{-d+4}{3}}r^{\frac{d+2}{3}}\ln r-\frac{3C_{1}%
}{d-1}r+C_{2}\left( \frac{r}{r_{1}}\right) ^{\frac{\left( d+2\right) }{3}} & 
, & A=\frac{1}{3} \\ 
-\frac{\left( d-1\right) \left( d+2\right) ^{2}}{d^{2}r_{1}^{\frac{d-2}{d+2}}%
}r^{\frac{3d+2}{d+2}}+\frac{Q^{2}\left( d+2\right) ^{2}}{d-1}\frac{r^{\frac{%
3d}{d+2}}}{r_{1}^{\frac{d-4}{d+2}}}+C_{1}r\ln r+C_{2}\frac{r}{r_{1}} & , & A=%
\frac{1}{d+2}%
\end{array}%
\right. .
\end{equation}%
The potential accordingly takes the form from (24) 
\begin{equation}
V=\left\{ 
\begin{array}{lcr}
\frac{r_{1}^{2A}}{r^{2A}}\frac{d\left( d-1\right) \left( A-1\right) }{\left(
d+2\right) \left( 2A-1\right) }+\frac{r_{1}^{4A}}{r^{4A}}\frac{\left(
1-A\right) d\left( d-1\right) Q^{2}}{\left( 3A-1\right) }-dA\left[ A\left(
d+2\right) -1\right] C_{2}\frac{r^{2\left( A-1\right) }}{r_{1}^{2A}} & , & 
A\neq \frac{1}{2},\frac{1}{3},\frac{1}{d+2} \\ 
\frac{d\left( d-1\right) Q^{2}r_{1}^{2}}{r^{2}}-\frac{d^{2}C_{2}}{4r_{1}r}+%
\frac{\left( d-1\right) r_{1}}{r}\left( d+2+d\ln r\right) & , & A=\frac{1}{2}
\\ 
2\left( d-1\right) d\left( \frac{r_{1}}{r}\right) ^{\frac{2}{3}}-\frac{1}{9}%
\frac{d\left( d-1\right) C_{2}}{r^{\frac{4}{3}}r_{1}^{\frac{2}{3}}}-\left( 
\frac{r_{1}}{r}\right) ^{\frac{4}{3}}\frac{dQ^{2}\left( 4\left( d-1\right)
\ln r+3d+9\right) }{3} & , & A=\frac{1}{3} \\ 
\left( d^{2}-1\right) \left( \frac{r_{1}}{r}\right) ^{\frac{2}{d+2}}-\left( 
\frac{r_{1}}{r}\right) ^{\frac{4}{d+2}}d\left( d+1\right) Q^{2}-\frac{r_{1}^{%
\frac{d}{d+2}}dC_{1}}{d+2}r^{-2\frac{d+1}{d+2}} & , & A=\frac{1}{d+2}%
\end{array}%
\right.
\end{equation}%
Let us note that $r_{1}$ is a constant introduced for dimensional reason and
without loss of generality we set it as $r_{1}=1$ in the sequel.

\subsection{\noindent Asymptotic Functions}

In this subsection we give the asymptotic behaviors of the general solution
found above. To do so we first rewrite the solution (30) in terms of $R$
(with $r_{1}=1$), 
\begin{equation}
u\left( R\right) =\left\{ 
\begin{array}{lcr}
\frac{\left( d-1\right) R^{\frac{2+A\left( d-2\right) }{A}}}{\left( A\left(
d-2\right) +1\right) \left( 2A-1\right) }-\frac{2\left( d-1\right) Q^{2}R^{%
\frac{A\left( d-4\right) +2}{A}}}{\left( A\left( d-4\right) +1\right) \left(
3A-1\right) }-\frac{C_{1}}{A\left( d+2\right) -1}R^{\frac{1}{A}}+C_{2}R^{d+2}
& , & A\neq \frac{1}{2},\frac{1}{3},\frac{1}{d+2} \\ 
-\frac{8\left( d-1\right) \ln (R)R^{d+2}}{d}-\frac{8\left( d-1\right)
Q^{2}R^{d}}{d-2}-\frac{2C_{1}}{d}R^{2}+C_{2}R^{d+2} & , & A=\frac{1}{2} \\ 
-\frac{9\left( d-1\right) R^{d+4}}{\left( d+1\right) }+36Q^{2}R^{d+2}\ln R-%
\frac{3C_{1}}{d-1}R^{3}+C_{2}R^{d+2} & , & A=\frac{1}{3} \\ 
-\frac{\left( d-1\right) \left( d+2\right) ^{2}}{d^{2}}R^{3d+2}+\frac{%
Q^{2}\left( d+2\right) ^{2}}{d-1}R^{3d}+C_{1}\left( d+2\right) R^{d+2}\ln
R+C_{2}R^{d+2} & , & A=\frac{1}{d+2}%
\end{array}%
\right. ,
\end{equation}%
and 
\begin{equation}
V\left( R\right) =\left\{ 
\begin{array}{lcr}
\frac{1}{R^{2}}\frac{d\left( d-1\right) \left( A-1\right) }{\left(
d+2\right) \left( 2A-1\right) }+\frac{1}{R^{4}}\frac{\left( 1-A\right)
d\left( d-1\right) Q^{2}}{\left( 3A-1\right) }-A\left( A\left( d+2\right)
-1\right) dC_{2}R^{\frac{2\left( A-1\right) }{A}} & , & A\neq \frac{1}{2},%
\frac{1}{3},\frac{1}{d+2} \\ 
\frac{d\left( d-1\right) Q^{2}}{R^{4}}-\frac{d^{2}C_{2}}{4R^{2}}+\frac{%
\left( d-1\right) }{R^{2}} & , & A=\frac{1}{2} \\ 
\frac{2\left( d-1\right) d}{R^{2}}-\frac{1}{9}\frac{d\left( d-1\right) C_{2}%
}{R^{4}}-\frac{dQ^{2}\left( d+3+\frac{4\left( d-1\right) }{9}\ln R\right) }{%
R^{4}} & , & A=\frac{1}{3} \\ 
\frac{\left( d^{2}-1\right) }{R^{2}}-\frac{d\left( d+1\right) Q^{2}}{R^{4}}-%
\frac{dC_{1}}{\left( d+2\right) R^{2\left( d+1\right) }} & , & A=\frac{1}{d+2%
}%
\end{array}%
\right. .
\end{equation}%
We also note that in terms of $R$ the line element (7) reads 
\begin{equation}
ds^{2}=-U\left( R\right) dt^{2}+\frac{R^{\frac{2\left( 1-A\right) }{A}}dR^{2}%
}{A^{2}U\left( R\right) }+R^{2}d\Omega _{d}^{2},
\end{equation}%
in which $U\left( R\right) =\frac{u}{R^{d}}$. For $0<A<\frac{1}{2}$ while $%
R\rightarrow \infty $ the line element becomes 
\begin{equation}
ds^{2}\simeq \xi ^{2}R^{\frac{2-2A}{A}}dt^{2}-\frac{dR^{2}}{A^{2}\xi ^{2}}%
+R^{2}d\Omega _{d}^{2}
\end{equation}%
in which $\xi ^{2}=-\frac{\left( d-1\right) }{\left( A\left( d-2\right)
+1\right) \left( 2A-1\right) }$ so that $R$ turns into time-like. For $\frac{%
1}{2}<A<1$, we obtain asymptotically 
\begin{equation}
ds^{2}\simeq -C_{2}R^{2}dt^{2}+\frac{R^{\frac{2\left( 1-2A\right) }{A}}dR^{2}%
}{A^{2}C_{2}}+R^{2}d\Omega _{d}^{2}.
\end{equation}%
It is observed that since $A\neq 1$ asymptotic form $(R\rightarrow \infty )$
excludes the case of AdS. For the case when $A=\frac{1}{2}$ one easily finds 
\begin{equation}
ds^{2}\simeq \lambda ^{2}\ln (R)R^{2}dt^{2}-\frac{4dR^{2}}{\lambda ^{2}\ln
(R)}+R^{2}d\Omega _{d}^{2}
\end{equation}%
where $\lambda ^{2}=\frac{8\left( d-1\right) }{d}$. It is needless to
restate that the roles of $t$ and $R$ change, i.e. $R$ becomes a time-like
coordinate. The asymptotic metrics (35) and (37) suggest also that we have
no Lifshitz {\cite{7}} asymptotes.

\section{Asymptotically Anti / de-Sitter Solutions for $d=4$}

In this section we consider a different ansatz for the scalar field which
reads%
\begin{equation}
\phi =\xi \ln \left( 1+\frac{r_{0}}{r}\right)
\end{equation}%
in which $\xi $ and $r_{0}>0$ are some real constants. We plug $\phi $ into
(14) to find 
\begin{equation}
R=\chi r\left( 1+\frac{r_{0}}{r}\right) ^{\beta +\frac{1}{2}}
\end{equation}%
in which $\chi $ is an integration constant and $\beta ^{2}=\frac{1}{4}-%
\frac{2}{d}\xi ^{2}.$ We note that due to this condition $\beta $ remains
bounded as%
\begin{equation}
-\frac{1}{2}\leq \beta \leq \frac{1}{2}
\end{equation}%
and consequently%
\begin{equation}
-\frac{1}{2\sqrt{2}}\leq \xi \leq \frac{1}{2\sqrt{2}}.
\end{equation}%
The general solution given for $u,$ unfortunately is not integrable in
closed form for a general dimension $d+2$ and for a general $\beta $. But
one can see that in $d+2$ dimension a specific choice for i.e., $\beta =-1/2$
yields 
\begin{equation}
U=\frac{1}{\chi ^{2}}-\frac{\left( d-1\right) Q^{2}}{\chi ^{4}r^{2}\left(
d-3\right) }-\frac{C_{1}}{r^{d-1}\chi ^{d}\left( d+1\right) }+\chi
^{2}C_{2}r^{2}
\end{equation}%
which in turn must give the general EYM solution for $d>3.$ This, of course,
implies $\chi =1,$ $C_{1}=m\left( d+1\right) $ and $C_{2}=-\frac{1}{3}%
\Lambda $ which casts the solution into%
\begin{equation}
U=1-\frac{\left( d-1\right) Q^{2}}{\left( d-3\right) r^{2}}-\frac{m}{r^{d-1}}%
-\frac{1}{3}\Lambda r^{2}.
\end{equation}%
Also the corresponding potential $V$ reads%
\begin{equation}
V=\frac{1}{3}d\left( d+1\right) \Lambda
\end{equation}%
which means that $\Lambda $ represents the cosmological constant.

Another case that we can integrate (23) is $d=4$ with $\beta =0$
(consequently $\xi ^{2}=\frac{1}{2}$) giving%
\begin{eqnarray}
U &=&\frac{r_{0}^{2}\left( 1+x\right) }{x^{2}}\Lambda -\frac{3\left(
2m-12Q^{2}r_{0}-r_{0}^{3}\right) \left( 1+x\right) \ln (1+x)}{x^{2}r_{0}^{3}}%
- \\
&&\frac{m\left( x+2\right) \left( x^{2}-6x-6\right) }{2xr_{0}^{3}\left(
1+x\right) }-\frac{6Q^{2}\left( 6+9x+2x^{2}\right) }{xr_{0}^{2}\left(
1+x\right) }-\frac{7x+6}{2x\left( 1+x\right) },  \notag
\end{eqnarray}%
for $x=\frac{r_{0}}{r}.$ We add also that to get the correct flat limit we
set $\chi =1$ and $C_{2}=\Lambda .$ Our other integration constant $C_{1}$
is related to the mass of the black hole solution so we set it as $m$. Here
it worths to see $\lim_{r\rightarrow 0/x\rightarrow \infty }U=-\infty $
which clearly implies that the solution admits an event horizon and
therefore the solution is a black hole.

Next, we find the form of the potential which is given by%
\begin{equation}
V=2\left[ \frac{6\Pi _{1}+\left( m-6Q^{2}r_{0}\right) \Pi _{2}}{r_{0}^{5}}%
\exp \left( -\sqrt{2}\phi \right) +2\Lambda \Pi _{3}\exp \left( -\frac{\phi 
}{\sqrt{2}}\right) +\frac{\Pi _{4}}{r_{0}^{2}}\exp \left( -\sqrt{2}\phi
\right) \right] ,
\end{equation}%
where%
\begin{equation*}
\Pi _{1}=\left( \exp \left( \frac{\phi }{\sqrt{2}}\right) \left( \exp \left( 
\frac{\phi }{\sqrt{2}}\right) +3\right) +1\right) \left(
-2m+12Q^{2}r_{0}+r_{0}^{3}\right) \left( \frac{\phi }{\sqrt{2}}\right) ,
\end{equation*}%
\begin{equation*}
\Pi _{2}=\left[ \exp \left( \sqrt{2}\phi \right) -1\right] \left[ \exp
\left( \frac{\phi }{\sqrt{2}}\right) \left( \exp \left( \frac{\phi }{\sqrt{2}%
}\right) +28\right) +1\right] ,
\end{equation*}%
\begin{equation*}
\Pi _{3}=\left[ \exp \left( \frac{\phi }{\sqrt{2}}\right) \left( \exp \left( 
\frac{\phi }{\sqrt{2}}\right) +3\right) +1\right] ,
\end{equation*}%
\begin{equation*}
\Pi _{4}=\left[ \exp \left( \frac{\phi }{\sqrt{2}}\right) -1\right] \left[
\exp \left( \frac{\phi }{\sqrt{2}}\right) \left( 16\exp \left( \frac{\phi }{%
\sqrt{2}}\right) +13\right) +1\right] .
\end{equation*}%
It is easy to observe that 
\begin{equation}
\lim_{\substack{ r\rightarrow \infty  \\ (x\rightarrow 0)}}V=20\Lambda ,
\end{equation}%
in which for zero cosmological constant it vanishes.

\section{Dynamics of Domain-Walls}

In this section we consider a $d+2-$dimensional bulk action supplemented by
surface terms {\cite{10}} 
\begin{equation}
S=\int_{\mathcal{M}}d^{d+2}x\sqrt{-g}\left[ \mathcal{R}-2\left( \partial
\phi \right) ^{2}-L_{\left( YM\right) }-V\left( \phi \right) \right]
+\int_{\Sigma }d^{d+1}x\sqrt{-h}\left\{ K\right\} +\int_{\Sigma }d^{d+1}x%
\sqrt{-h}L_{\left( DW\right) },
\end{equation}%
where the DW Lagrangian will be given by $L_{\left( DW\right) }=-\hat{V}%
\left( \phi \right) $ as the induced potential on the DW. $\left\{ K\right\} 
$ stands for the trace of the extrinsic curvature tensor $K_{ij}$ of DW with
the induced metric $h_{ij}$ ($h=$ $\left\vert g_{ij}\right\vert $).

Herein $\Sigma $ is the $(d+1)-$dimensional DW in a $(d+2)-$dimensional bulk 
$\mathcal{M}$ which splits the background bulk into two $(d+2)-$dimensional
spacetimes $\mathcal{M}_{\pm }$. Note that $\pm $ refer to the normal
directions on the DW. The metric we shall work on is given by (7) whose
parameters are chosen such that $\lim_{r\rightarrow \infty }U\left( r\right)
=\infty $ and for the case of BH $r_{h}<r=a$. For the non-BH case we make
the choice $0<r=a$. Let's impose now the following condition 
\begin{equation}
-U\left( a\right) \left( \frac{dt}{d\tau }\right) ^{2}+\frac{1}{U\left(
a\right) }\left( \frac{da}{d\tau }\right) ^{2}=-1
\end{equation}%
in which the location of the DW is given by $r=a\left( \tau \right) $, with
the proper time $\tau $. Therefore the line element on the DW becomes 
\begin{equation}
ds_{dw}^{2}=-d\tau ^{2}+a\left( \tau \right) ^{2}d\Omega _{d}^{2}.
\end{equation}%
which is a FRW metric in $(d+1)-$dimensions with the radius function $%
a\left( \tau \right) $. Imposing the boundary conditions i.e., the
Darmois-Israel conditions on the wall {\cite{4}} leads to 
\begin{equation}
-\left( \left\langle K_{i}^{j}\right\rangle -\left\langle K\right\rangle
\delta _{i}^{j}\right) =S_{i}^{j},
\end{equation}%
where 
\begin{equation}
S_{ij}=\frac{1}{\sqrt{-h}}\frac{2\delta }{\delta g^{ij}}\int d^{d+1}x\sqrt{-h%
}\left( -\hat{V}\left( \phi \right) \right) ,
\end{equation}%
is the surface energy-momentum tensor on the DW. Note that a bracket ($%
\left\langle .\right\rangle $) denotes a jump across $\Sigma $. The latter
yields 
\begin{equation}
S_{i}^{j}=-\hat{V}\left( \phi \right) \delta _{i}^{j}.
\end{equation}%
which after (50) and (52), gives the energy density $\sigma $ and the
surface pressures $p_{\theta _{i}}$ for generic metric functions $f\left(
r\right) $ and $R\left( r\right) $ with $r=a\left( \tau \right) $. The
overall results are given by 
\begin{equation}
\sigma =-S_{\tau }^{\tau }=-\frac{d}{4\pi }\left( \sqrt{U\left( a\right) +%
\dot{a}^{2}}\frac{R^{\prime }}{R}\right)
\end{equation}%
\begin{equation}
S_{\theta _{i}}^{\theta _{i}}=p_{\theta _{i}}=\frac{1}{8\pi }\left( \frac{%
U^{\prime }+2\ddot{a}}{\sqrt{U\left( a\right) +\dot{a}^{2}}}+2\left(
d-1\right) \sqrt{U\left( a\right) +\dot{a}^{2}}\frac{R^{\prime }}{R}\right)
\end{equation}%
in which a dot "$\cdot $" and prime "$^{\prime }$" means $\frac{d}{d\tau }$
and $\frac{d}{da}$, respectively. The junction conditions, therefore read 
\begin{equation}
\frac{d}{4\pi }\left( \sqrt{U\left( a\right) +\dot{a}^{2}}\frac{R^{\prime }}{%
R}\right) =-\hat{V}\left( \phi \right)
\end{equation}%
\begin{equation}
\frac{1}{8\pi }\left( \frac{U^{\prime }+2\ddot{a}}{\sqrt{U\left( a\right) +%
\dot{a}^{2}}}+2\left( d-1\right) \sqrt{U\left( a\right) +\dot{a}^{2}}\frac{%
R^{\prime }}{R}\right) =-\hat{V}\left( \phi \right) ,
\end{equation}%
which upon using 
\begin{equation}
\frac{U^{\prime }+2\ddot{a}}{\sqrt{U\left( a\right) +\dot{a}^{2}}}=\frac{2}{%
\dot{a}}\frac{d}{d\tau }\left( \sqrt{U\left( a\right) +\dot{a}^{2}}\right)
\end{equation}%
and Eq. (57) casts into 
\begin{equation}
\frac{U^{\prime }+2\ddot{a}}{\sqrt{U\left( a\right) +\dot{a}^{2}}}=\frac{2}{%
\dot{a}}\frac{d}{d\tau }\left( -\frac{4\pi }{d}\hat{V}\left( \phi \right) 
\frac{R}{R^{\prime }}\right) =\frac{d}{da}\left( -\frac{8\pi }{d}\hat{V}%
\left( \phi \right) \frac{R}{R^{\prime }}\right) .
\end{equation}%
Finally we find 
\begin{equation}
\frac{d}{da}\left( \hat{V}\left( \phi \right) \frac{R}{R^{\prime }}\right) =%
\hat{V}\left( \phi \right) .
\end{equation}%
This equation admits a simple relation between $R(r)$ and $\hat{V}\left(
\phi \right) $ given by 
\begin{equation}
R^{\prime }(r)=\xi \hat{V}\left( \phi \right)
\end{equation}%
with $\xi =$ constant. Using above with some manipulation we obtain the
one-dimensional equation 
\begin{equation}
\dot{a}^{2}+V_{eff}\left( a\right) =0
\end{equation}%
where the effective potential is defined by 
\begin{equation}
V_{eff}\left( a\right) =U\left( a\right) -\left( \frac{4\pi R}{d\xi }\right)
^{2}.
\end{equation}%
In order that (62) admits a solution as the radius on the DW universe we
must have $V_{eff}\left( a\right) <0$, which naturally restricts the
probable forms of the potential function $V\left( \phi \right) $. Further,
equations (61) yields 
\begin{equation}
\hat{V}\left( \phi \right) =\frac{Ar^{A-1}}{\xi }\text{,}
\end{equation}%
The following boundary condition for the scalar field (see Eq. (38) in
reference {\cite{10}}) 
\begin{equation}
\frac{\partial \phi }{\partial R}=-\frac{d}{R}\frac{1}{\hat{V}\left( \phi
\right) }\frac{\partial \hat{V}\left( \phi \right) }{\partial \phi }
\end{equation}%
holds \ automatically (after fixing $\xi =2A$ for convenience). Based on Eq.
(63) we plot $V_{eff}\left( a\right) $ for the cases of a non-black hole
(Fig. 1a) and an extremal black hole (Fig. 1b) as examples to verify double
bounces.

\section{Conclusion}

Einstein-Yang-Mills (EYM) fields, minimally coupled with a massless scalar
field $\phi $ supplemented by a scalar potential $V\left( \phi \right) $ are
considered. The EYMS system admits exact, both black hole and non-black hole
solutions. Depending on the scalar field the potential $V\left( \phi \right) 
$ has a large spectrum of possible values with the YM-field. The novelty
with the inclusion of the potential $V\left( \phi \right) $ has significant
contributions. Firstly, if the potential $V\left( \phi \right) $ is set to
zero the scalar field becomes trivial. That is, the YM field within
Einstein's general relativity at least in the assumed ansatz spacetime can't
coexist with a minimally coupled scalar field. The non-asymptotically flat
and non-asymptotically AdS black hole solutions obtained by our formalism
with a scalar hair. Depending on the dimensionality of the spacetime and the
scalar field ansatz the picture may change entirely. This we show with an
explicit example in $6-$dimensional spacetime which admits an anti de Sitter
asymptote. Secondly, the richness brought in by the additional potential $%
V\left( \phi \right) $ (i.e. Section III)in the bulk spacetime of dimension (%
$d+2$) is employed in the construction of a Domain Wall (DW) universe as a
brane in ($d+1$)-dimensions. Specifically, we are interested to construct a
DW brane as a FRW universe which admits both a minimum and a maximum
bounces. We have shown that this is possible, so that once our world is such
a brane it will oscillate between these limits ad infinitum (see Fig. 1).
Finally, since a constant potential amounts to a cosmological term a
space-filling uniform scalar field may be the cause of the cosmological
constant. \ \ \

\textbf{Figure Caption:}

Examples of Einstein-Yang-Mills-Scalar (EYMS) solutions with double bounces
on the domain wall. For technical reasons we restrict ourselves to $d=5$
alone. Fig. 1a is the case of non-black hole, while Fig.1b refers to the
case of an extremal black hole. The effective potential $V_{eff}\left(
a\right) $ is given in Eq. (63). From Eq. (62) the universe admits only the
possibility of $V_{eff}\left( a\right) <0$. This gives rise to an
oscillatory FRW universe on the DW for such an EYMS system when supplemented
by a scalar potential in the Lagrangian.

\bigskip

\end{document}